\documentclass{article}

\usepackage{arxiv}

\usepackage[utf8]{inputenc} 
\usepackage[T1]{fontenc}    
\usepackage{hyperref}       
\usepackage{url}            
\usepackage{booktabs}       
\usepackage{amsfonts}       
\usepackage{nicefrac}       
\usepackage{microtype}      
\usepackage{lipsum}
\usepackage{graphicx}
\graphicspath{ {./images/} }

\title{A comparative analysis of CT degradation for LDCT nodule classification using radiomics}

\author{
 Jiaying Liu \\
  Department of Electronics, Informatics \\
   and Bioengineering, Politecnico di Milano\\
  Milano, Italy\\
  \texttt{jiaying.liu@polimi.it} \\
   \And
 Anna Corti \\
  Department of Electronics, Informatics \\
  and Bioengineering, Politecnico di Milano\\
  Milano, Italy\\
  \texttt{anna.corti@polimi.it} \\
  \And
 Valentina D.A. Corino\\
  Department of Electronics, Informatics \\
  and Bioengineering, Politecnico di Milano\\
  Cardiotech Lab, Centro Cardiologico Monzino IRCCS\\
  Milano, Italy\\
  \texttt{valentina.corino@polimi.it} \\
  \And
 Luca Mainardi\\
  Department of Electronics, Informatics \\
  and Bioengineering, Politecnico di Milano\\
  Milano, Italy\\
  \texttt{luca.mainardi@polimi.it} \\
}

\begin{document}
\maketitle
\begin{abstract}
Low-dose computed tomography (LDCT) is the standard modality for lung cancer screening, known for its low radiation dose but high noise levels. While existing literature focuses on denoising LDCT images, comparative research on simulating LDCT characteristics to directly use these images for model development is lacking. This study shifts the focus from denoising images to degrading available standard-dose CT (SDCT) data, generating synthetic images for data augmentation to train classifiers for screening-detected nodules.
We compare three degradation methods: (1) a sinogram domain statistical noise insertion; (2) replicate a validated physics-based simulation using Pix2Pix; and (3) unpaired CycleGAN. The generated images were utilized to simulate LDCT screening scenario replacing 695 SDCT cases from the LIDC-IDRI dataset, from which radiomic features were extracted to train machine learning models for lung nodule classification.
Regarding image quality, CycleGAN achieved the best Fréchet inception distance (0.1734) and kernel inception distance ($0.0813 \pm 0.1002$) scores, indicating distributional alignment with the target low-dose domain. In the nodule classification task, results confirmed the necessity of domain adaptation since a baseline model trained on non-degraded SDCT data failed to generalize to the real LDCT set (AUC 0.789) with a low sensitivity (0.571). Degraded images generated using CycleGAN approach led to the most balanced performance on the classification task using Adam Booster classifier, achieving an AUC of 0.861, sensitivity of 0.743 and specificity of 0.858 in the independent test.
Our findings confirm that generating synthetic LDCT data from standard-dose scans is a viable strategy for training robust nodule classifiers for screening detected nodules.
\end{abstract}


\section{Introduction}
Lung cancer remains the leading cause of cancer-related mortality worldwide \cite{Sung2021}. Early detection through lung cancer screening has been shown to significantly reduce mortality as evidenced in several clinical trials \cite{deKoning2020, NLST2011}. Through these trials, low-dose computed tomography (LDCT) has emerged as the preferred imaging modality for this screening, as it provides the necessary diagnostic capability while minimizing the inherent risk of cumulative radiation exposure associated with standard-dose computed tomography (SDCT) scans (also referred to as routine-dose and full-dose). Nonetheless, lower radiation dose results in a lower signal-to-noise ratio and in images with lower contrast, blurring and increased noise and artifacts, all of which may affect clinicians’ diagnostic results. Consequently, a substantial portion of the current literature focuses on LDCT image restoration methods to improve LDCT image quality \cite{Kim2024}. Conventional restoration methods can be categorized into three groups: sinogram domain filtering, iterative reconstruction and image domain restoration \cite{Kulathilake2023}. The sinogram domain filtering directly works on the raw projection data before back-projection, allowing for accurate computation of noise statistics. Iterative reconstruction methods alternate between the sinogram and image domains, which results in high computational expense. Image domain restoration is considered a post-processing method directly applied to reconstructed images. These conventional methods either requires access to raw projection data, which is scanner-specific and inaccessible, or paired SDCT-LDCT image datasets for training, which is problematic as it is unethical to expose patients to repeated computed tomography (CT) scans. To overcome these limitations, some researchers create synthetic LDCT images by degrading existing SDCT data to train denoising \cite{Zhao2019, Gholizadeh2019, Chen2017}. Others rely on the dataset from the 2016 LDCT Grand challenge, yet even this standard benchmark was generated using a validated noise insertion method rather than raw clinical paired \cite{Eulig2024,McCollough2017}.  Therefore, as paired clinical data is not publicly available, the field of denoising relies on synthetic approaches.  While these approaches could be effective in benchmarks, there are two drawbacks: (1) the noise simulation is often treated as a secondary step, potentially compromising the fidelity of the synthetic images and the real-world generalizability of models trained on them; and (2) the cycles of synthetic degradation and subsequent denoising are computationally expensive and time-consuming during both development and clinical implementation. 

This study shifts the focus from denoising LDCT to generating diagnostically representative LDCT images from existing SDCT data, by a degradation step. Instead of denoising the images, this study aims to (1) compare degradation methods for synthetic image generation and (2) use these generated images to train and validate a robust classifier using radiomics. 

Radiomics is an approach that extracts high-throughput quantitative features from a region of interest, allowing for the acquisition of more accurate information that may be hidden from the human eye for analysis and predictive modeling in precision medicine. This approach faces a practical limitation: the lack of annotated LDCT datasets. To overcome this, we leverage the Lung Image Database Consortium and Image Database Resource Initiative (LIDC-IDRI) dataset, which contains mainly SDCT images but also LDCT, and propose a framework to generate simulated LDCT scans to bridge the domain gap for model training, reserving the real LDCT scans for model validation and testing. 

This paper aims to: (1) provide a systematic comparison and evaluation of different LDCT degradation methods; and (2) introduce an approach that bypasses denoising and instead validates the feasibility of using SDCT images for developing downstream classification models.

\section{Materials and Methods}
\label{sec:headings}
\subsection{Datasets}
This study uses two datasets: the LIDC-IDRI \cite{Armato2011} and the Low-Dose CT Image and Projection Data \cite{Moen2021}, which expands upon the subset originally released for the 2016 LDCT Grand Challenge.

\begin{figure} 
    \centering
    \includegraphics[width=0.8\linewidth]{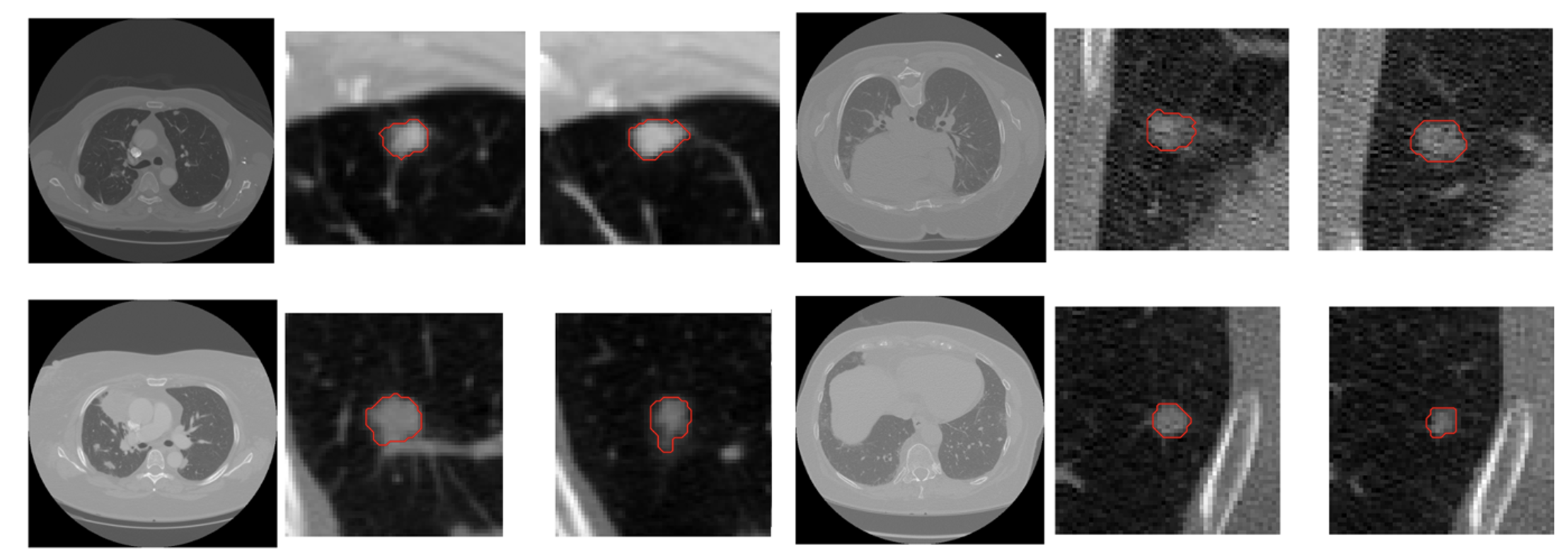}
    \caption{Examples of SDCT (a) and LDCT (b) subjects from the LIDC-IDRI dataset and examples of the corresponding nodule annotations.}
  \label{fig:fig1}
\end{figure}

\paragraph{LIDC-IDRI dataset.}
The LIDC-IDRI dataset consists of 1018 thoracic CT scans acquired from 1010 participants, with nodule annotations performed by one to four experienced radiologists. Consistent with the literature \cite{Snowsill2018,Rampinelli2013}, we defined LDCT as acquisitions with a tube current $\le 80 mA$, equivalent to 40 $mAs$ assuming a typical gantry rotation time of 0.5 seconds. Fig. \ref{fig:fig1} shows examples of SDCT (a) and LDCT (b) subjects from the LIDC-IDRI dataset and examples of the corresponding nodule annotations.

All images were preprocessed to reduce the image-related sources of variability. The preprocessing steps included Hounsfield Unit (HU) conversion, intensity range clipping to a window of [-1200,600] HU, denoising using a 3D Gaussian filter with a $3 \times 3 \times 3$ voxel kernel and $\sigma = 0.5$, and image normalization to the [0,1] range. 

After excluding the duplicated participants, a total of 2625 nodules, each assigned as $\ge 3 mm$ (lesions with the greatest inplane dimension in the range 3–30 mm) marked by at least one radiologist, were included in this study.  Table \ref{tab:lidc} describes the LIDC-IDRI dataset.

\begin{table}[htbp]
    \centering
    \caption{Summary of the LIDC-IDRI dataset used in this study.}
    \label{tab:lidc}
    \begin{tabular}{lcc}
        \toprule
        LIDC-IDRI dataset & \textbf{SDCT} & \textbf{LDCT} \\
        \midrule
        \textbf{Nr. participants}                    & 695  & 315 \\
        \textbf{Nr. nodules}                         & 1950 & 675 \\
        \textbf{Non-malignant (average likelihood <4)} & 1725 & 606 \\
        \textbf{Malignant (average likelihood >4)}   & 225  & 69  \\
        \bottomrule
    \end{tabular}
\end{table}

\begin{figure} 
    \centering
    \includegraphics[width=0.8\linewidth]{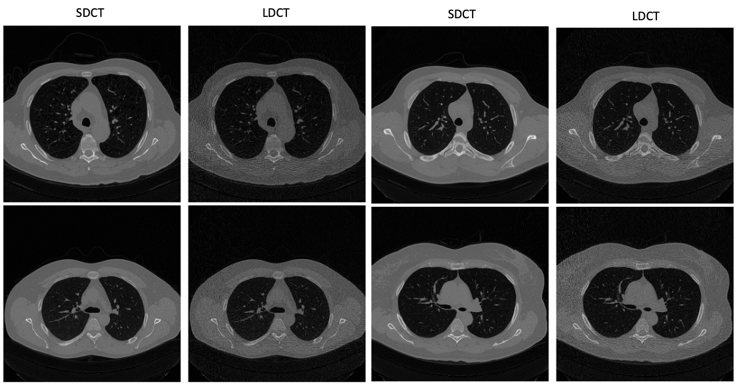}
    \caption{Example of paired SDCT and simulated LDCT slices from the Low-dose CT image and projection dataset.}
  \label{fig:fig2}
\end{figure}

\paragraph{Low-dose CT image and projection dataset.}
The Low-Dose CT Image and Projection dataset \cite{Moen2021} is composed of full-dose and quarter-dose projection data and images from 299 patients, including CT scans of the head, chest, and abdomen. For our study, we focused on a subset of 50 paired chest CT scans, which included both SDCT and simulated LDCT images. Fig.~\ref{fig:fig2} shows an example of paired SDCT and simulated LDCT slices from the dataset. Reduced dose projection data were simulated for each scan using a validated noise-insertion method that considers the effects of the bowtie filter, automatic tube current modulation and electronic noise \cite{Yu2012}. 

\subsection{Methods}

LDCT images are degraded by quantum noise and various artifacts during the image acquisition, which lead to a non-uniform distribution of noise over the image space as well as the presence of blurring and streaking artifacts (see Fig.\ref{fig:fig1} and Fig.\ref{fig:fig2}). While typical LDCT simulation algorithms inject Poisson noise into SDCT sinograms to mimic the dominant quantum noise, this approach oversimplifies the complex, stochastic nature of real LDCT noise. Therefore, to more accurately simulate LDCT images using SDCT data, we compared three degradation techniques: 1) sinogram domain noise injection \cite{Zeng2015}, 2) a validated physics-based degradation model using Pix2Pix \cite{Yu2012, Isola2017}, and 3) cycle-consistent generative adversarial network (CycleGAN), a popular unpaired image-to-image translation method \cite{Zhu2017}. A detailed description of these methods and the implemented training strategies is provided in the following sections.

\begin{figure} 
    \centering
    \includegraphics[width=0.8\linewidth]{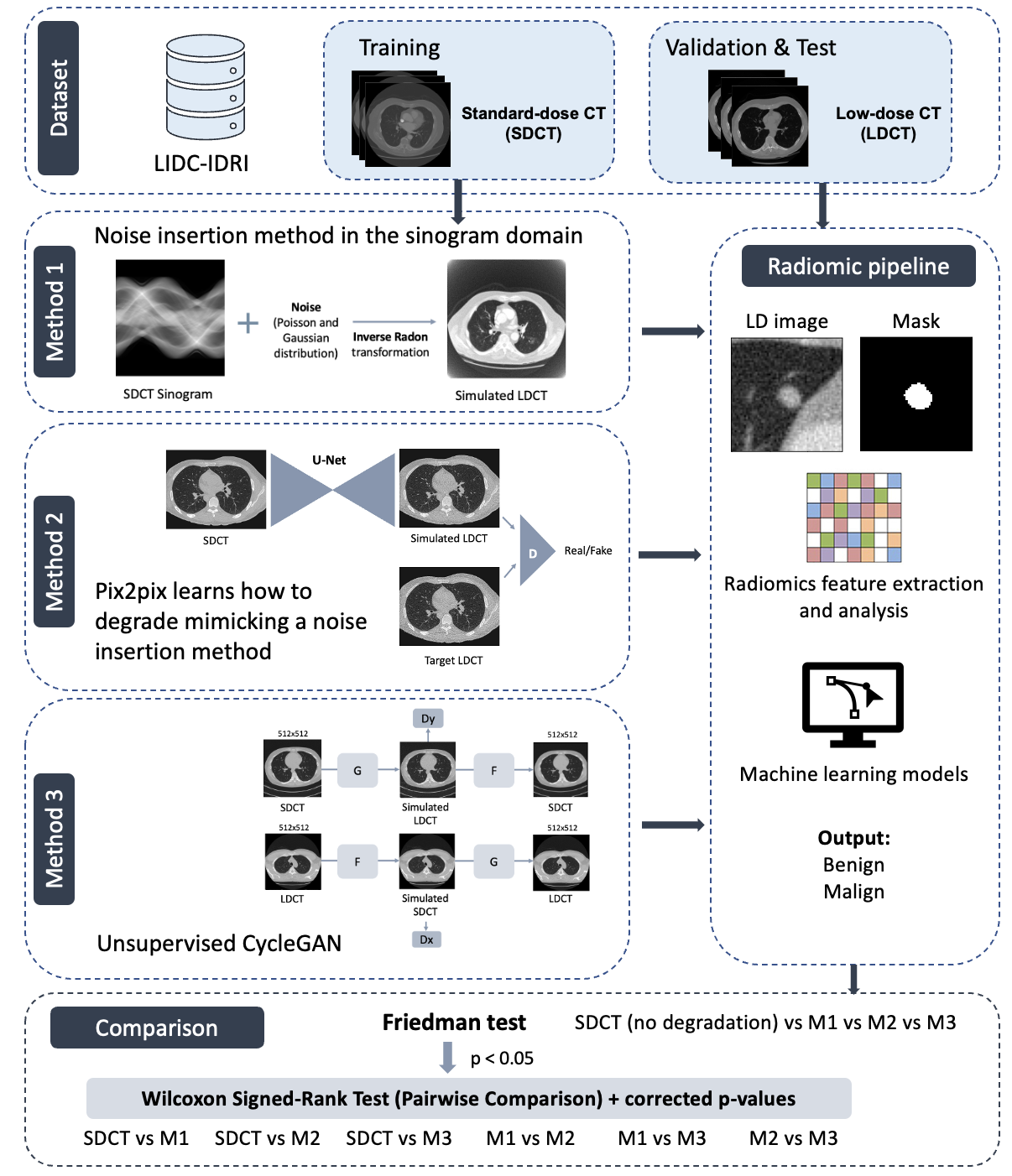}
    \caption{Overview of the pipeline.}
  \label{fig:fig3_pipeline}
\end{figure}

To compare the LDCT simulation methods, we propose a pipeline (Fig.\ref{fig:fig3_pipeline}) that includes a radiomics-based nodule classification task. First, the LIDC-IDRI dataset was split into SDCT, which is reserved for training, and LDCT, which is used for 50\% validation and 50\% testing. Then, the SDCT training set is degraded following the three selected degradation methods. For each degradation method, the pipeline proceeds to extract radiomic features from nodule annotations, perform feature selection, train and test machine learning models.

The pipeline was developed using Python 3.9. The code can be found at \url{https://github.com/jiayingliu423/LDCT-degradation-and-nodule-classification}.

\subsubsection{Method 1 – Simple sinogram-domain simulation}

\begin{figure} 
    \centering
    \includegraphics[width=0.6\linewidth]{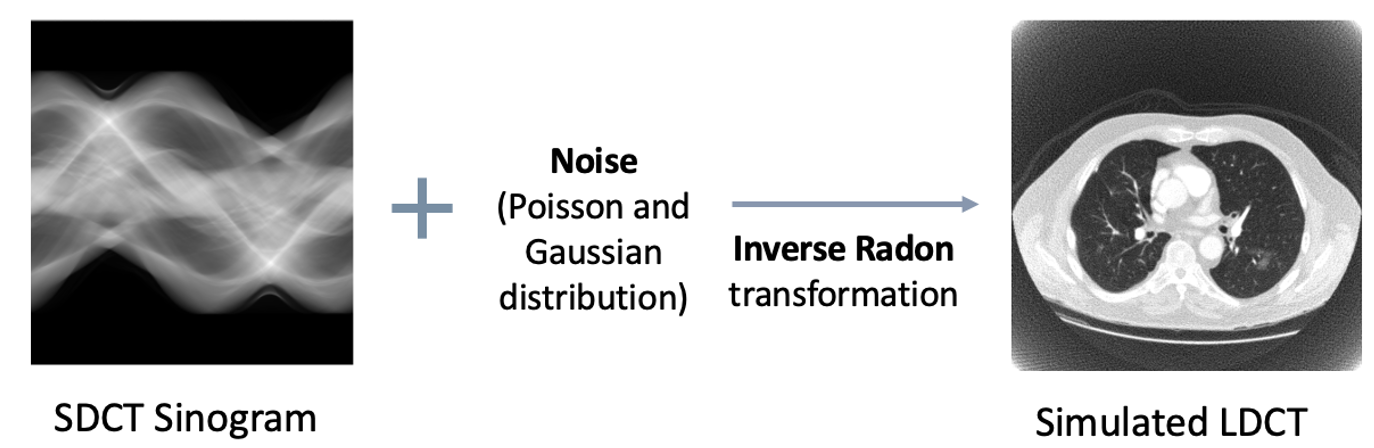}
    \caption{Pipeline to simulate low-dose computed tomography (LDCT). The Radon transformation is applied to the standard-dose computed tomography (SDCT) image, then electronic and quantum noise effects are included by noise injection. The final step involves inverse Radon transformation to reconstruct the LDCT image. Adapted from \cite{Liu2024}.}
  \label{fig:fig4}
\end{figure}

The first degradation method (\textbf{method 1}) is a simple low-dose CT simulation strategy in the sinogram domain using the raw data from a high-dose scan as proposed by \cite{Zeng2015} and utilized in the work of \cite{Liu2024}. This method simulates LDCT by adding Poisson and Gaussian noise to mimic the effect of electronic and quantum noise and has the advantage of being computationally lightweight while requiring minimal scanner-specific parameters. Briefly, standard dose sinogram data were obtained by using Radon transformation; subsequently, simulated Poisson and Gaussian noise were added so that the low dose sinogram data were generated as shown in Fig. \ref{fig:fig4}. Equation \ref{eq:eq_1} describes the calculation for the low dose sinogram data ($p_{ld,sim}$):
\begin{equation}
    p_{ld,sim} = \log \frac{I^{0}_{ld,sim}}{
        \mathrm{Poisson}\!\left(I^{0}_{ld,sim}(s) \cdot \exp(-p_{sd})\right)
        + \mathrm{Gaussian}(m_e,\, \sigma_e^{2})
    }
    \label{eq:eq_1}
\end{equation}

where $I_{ld,sim}^0$ is the simulated low-dose scan incident flux, $p_{sd}$ is the standard-dose sinogram data obtained from the original SDCT images using Radon transformation, $m_e$ and $\sigma_e^2$ are the mean and variance of the electronic noise, respectively. Finally, the simulated LDCT images were reconstructed from the low-dose sinogram data using the inverse Radon transform. 

\subsubsection{Method 2 – Physics-based degradation model replication using Pix2Pix}

The second degradation method (\textbf{method 2}) studied in this paper is inspired by the LDCT grand challenge, which uses a validated physics-based noise modeling technique from Yu et al. \cite{Yu2012} to generate the low-dose images present in the challenge dataset. Yu et al.’s technique incorporates the effect of bowtie filters, automatic exposure control and electronic noise. The noise insertion method requires scaling the incident photon number and adding normally distributed stochastic noise, as described in their equation: 
\begin{equation}
    \widetilde{P_B}\approx P_A+\sqrt{\frac{1-a}{a}\cdot\frac{\exp{\left(P_A\right)}}{N_{0A}}\cdot(1+\frac{1+a}{a}\cdot\frac{N_e\cdot\exp{\left(P_A\right)}}{N_{0A}})}x      \label{eq_2}
\end{equation}
where $\widetilde{P_B}$ is the simulated low dose data, $P_A$ is data at a higher tube current setting, $a (0<a<1)$ is a dose scaling factor, $N_{0A}$ is the incident photon flux corresponding to $P_A$,  $x$  is a normally distributed stochastic process with zero mean and a unit variance, $N_e$ represents the electronic noise.

Directly reproducing this model requires access to raw projection data, incident flux calibration and detailed scanner-specific variables, which are often inaccessible. We, therefore, leveraged the publicly available LDCT image and projection data dataset (challenge dataset)~\cite{Moen2021}, which was generated using the \cite{Yu2012} model and we estimate the scanner specific SDCT-to-LDCT transformation using the Pix2Pix \cite{Isola2017}.

\begin{figure} 
    \centering
    \includegraphics[width=0.6\linewidth]{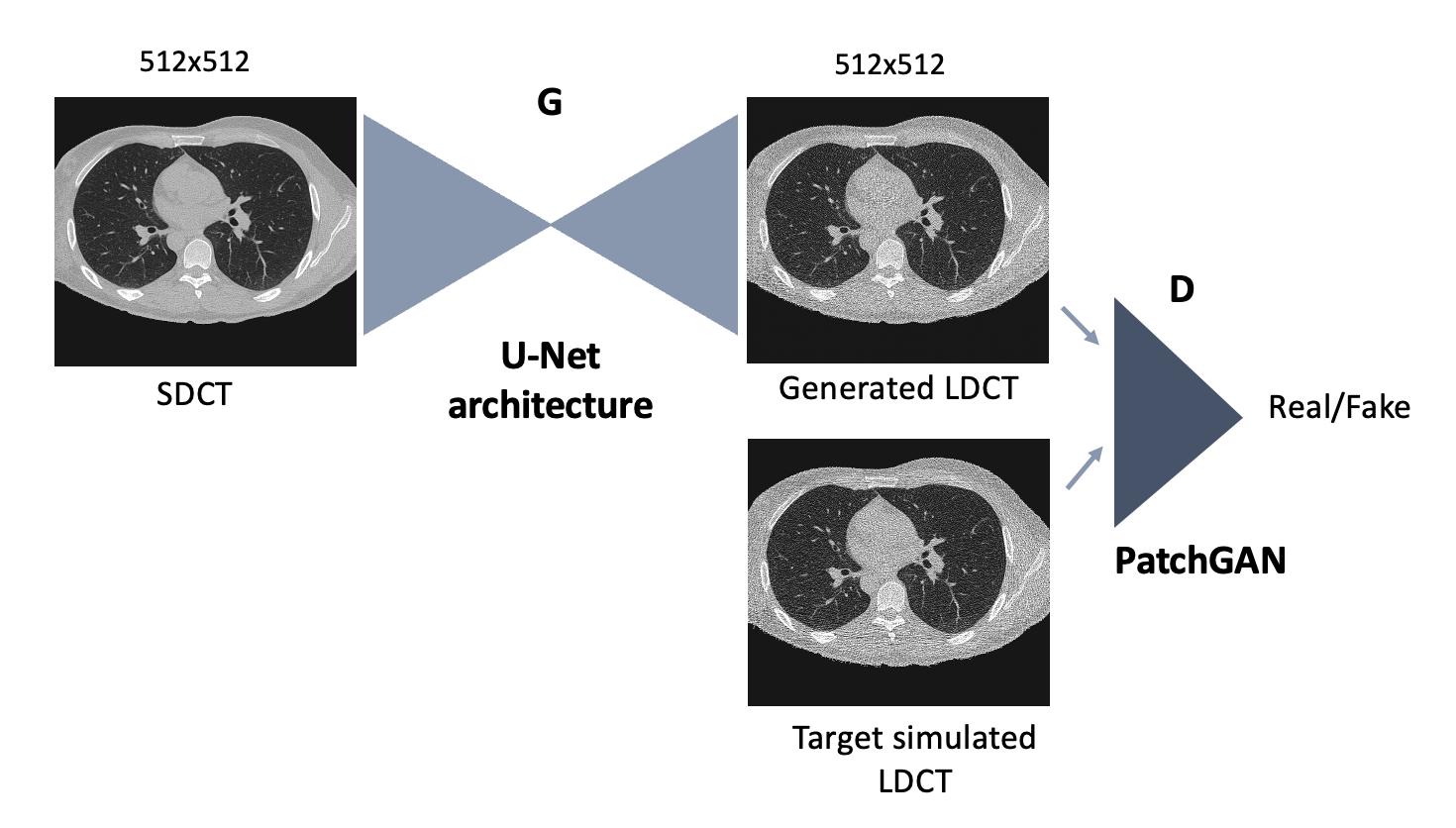}
    \caption{Architecture of Pix2Pix, composed by a generator (G) and a discriminator (D), used to replicate the degradation method from \cite{Yu2012}, leveraging paired images provided by the low-dose grand challenge dataset. SDCT: standard-dose CT, LDCT: low-dose CT.}
  \label{fig:fig5}
\end{figure}

Pix2pix, a conditional generative adversarial network (GAN), was selected for its ability to utilize paired training data to enforce strict anatomical consistency while synthesizing noise textures. The framework, as shown in Fig.\ref{fig:fig5}, consists of 1) a generator (G), composed by U-Net architecture with skip connections, and 2) a discriminator (D) PatchGAN that classified local NxN image patches as real or fake. In this way, the generator with the help of the discriminator which focuses on local patches, is forced to learn to simulate low dose images noise characteristics. The original objective function minimizes a combination of (a) the conditional adversarial loss ($L_{cGAN}$), which encourages the generator to produce realistic low-dose images that are indistinguishable from the low-dose domain; and (b) the $L1$ reconstruction loss ($L_{L1}$) which enforces pixel-wise similarity between the generated low-dose image output and the paired ground truth image. We included the structural similarity index (SSIM) loss term in addition to the standard adversarial and $L1$ loss to further enhance the perceptual quality and structural preservation of the CT images. The loss function is defined as:
\begin{equation}
    L\left(G,D\right)=\ L_{cGAN}\left(G,D\right)+\lambda_{L1}L_{L1}\left(G\right)+\lambda_{SSIM}L_{SSIM}(G)
    \label{eq_3}
\end{equation}

where $\lambda_{L1}$ and $\lambda_{SSIM}$ are the weights for the correspondent losses. In our experiment, we set both $\lambda_{L1}=50$  and $\lambda_{SSIM}=50$.

\paragraph{Training strategy}

The 50 paired chest CT scans, with consistent 512x512 pixels per slice, from the low-dose grand challenge dataset was partitioned into an 80-20 split for training and validation, respectively. The model was trained for 200 epochs (100 epochs at a constant learning followed by 100 epochs of linear decay) with a batch size of 32. Evaluation was performed using multi-scale SSIM (MS-SSIM) and mean absolute error (MAE) to assess structural accuracy.

\subsubsection{Method 3 – Unpaired image generation using CycleGAN }

\begin{figure} 
    \centering
    \includegraphics[width=0.6\linewidth]{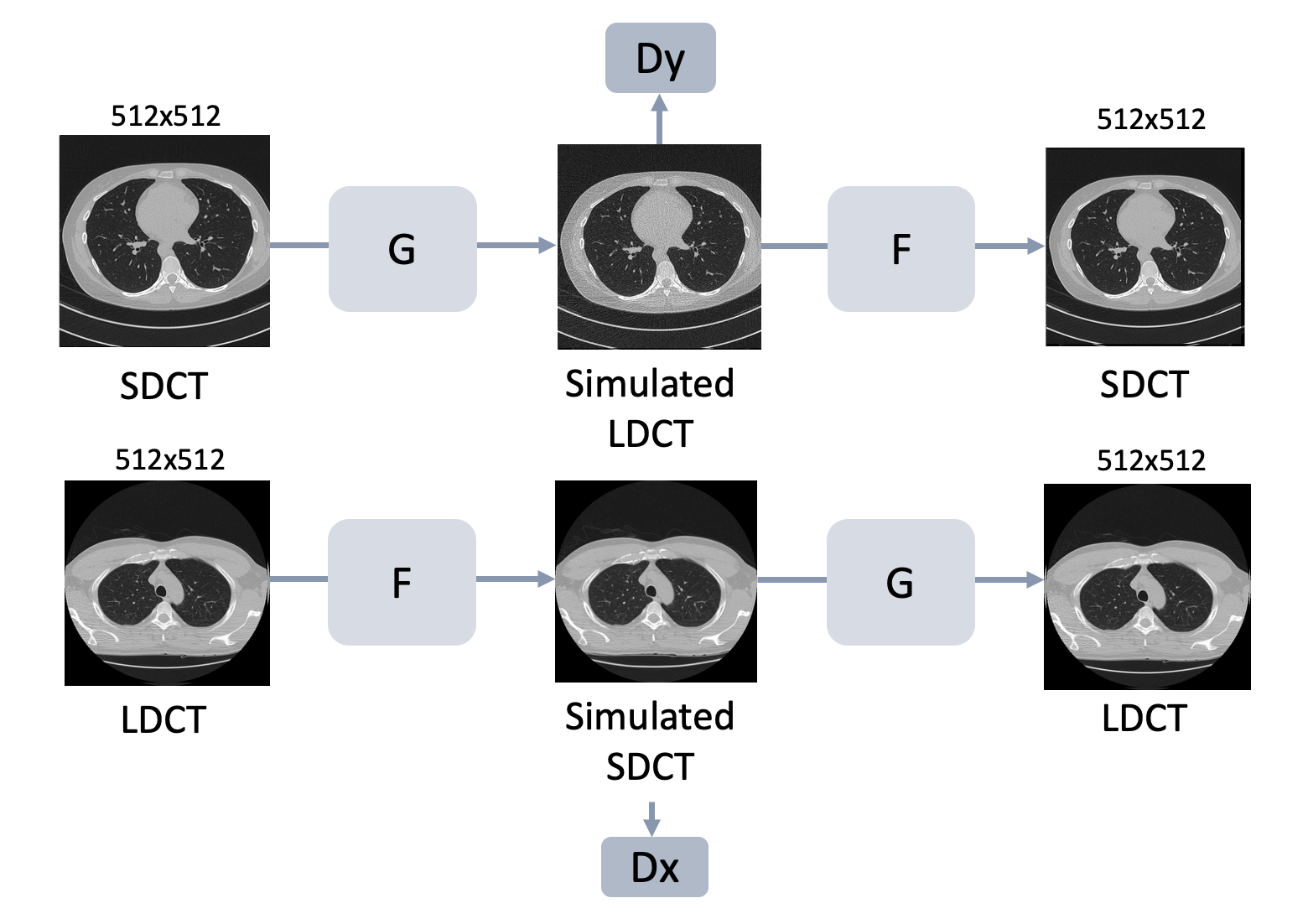}
    \caption{Schematic representation of the CycleGAN model. Generator G translates images from the SDCT to the LDCT, and F performs the reverse translation from LDCT to SDCT. Discriminator $D_y$ aims to distinguish between real LDCT images and generated LDCT images ($G(x)$) and $D_x$ should differentiate between real SDCT images and generated SDCT images $F(x)$). }
  \label{fig:fig6}
\end{figure}

For the third degradation method (\textbf{method 3}), we utilize CycleGAN~\cite{Zhu2017}, a popular GAN designed for unpaired image-to-image translation, capable of learning the characteristic noise distribution and artifacts of the LDCT domain directly from unpaired SDCT and LDCT images. CycleGAN, as shown in Fig.~\ref{fig:fig6}, is composed of two mapping functions (generators): $G:X\rightarrow Y$ and $F:Y\rightarrow X$; and two associated adversarial discriminators: $D_x$ and $D_y$. In our study, generator G translates images from the SDCT (domain X) to the LDCT (domain Y), and F performs the reverse translation from LDCT to SDCT. The discriminator $D_y$ aims to distinguish between real LDCT images and generated LDCT images ($G(x)$). In this way, $D_y$ encourages generator G to produce realistic LDCT images that are indistinguishable from the true LDCT images. Similarly, $D_x$ should differentiate between real SDCT images and generated SDCT images $F(x)$, guiding generator F to create realist SDCT images. 

The overall objective function minimizes a combination of a) the adversarial loss ($\mathcal{L}_{GAN}$), which encourages the generated images to match the distribution of images in the target domain (low-dose); and b) cycle consistency loss ($\mathcal{L}_{cyc}$) which enforces the intuition that an image translated from $X$ to $Y$ and back to $X$ (and vice versa) should be close to the original image, i.e., $F\left(G\left(x\right)\right)\approx x$ and $G\left(F\left(y\right)\right)\approx y$. The complete loss function is defined as:

\begin{equation}
    {{\mathcal{L}\left(G,F,D_X,D_Y\right)=\ \mathcal{L}}_{GAN}\ \left(G,\ D_Y,\ X,Y\right)+\mathcal{L}_{GAN}\ \left(F,\ D_X,\ Y,X\right)+\lambda\mathcal{L}}_{cyc}(G,F)
    \label{eq_4}
\end{equation}
where $\lambda$ controls the relative importance of the two losses. 

\paragraph{Training strategy}
The SDCT images were split into 80-20 for training and testing (see Table \ref{tab:table_2}), resulting in 556 subjects and 139,301 2D images in the training set and 139 subjects and 35,280 2D images for testing. All the LDCT images, consisting of 315 subjects and 68,302 2D images, were used for training. The input and output image sizes were 512x512. For every 5, 10, 15, 20 and 25 epochs, we used the reserved SDCT test set to evaluate the CycleGAN model using Fréchet Inception Distance (FID) and Kernel Inception Distance (KID) metrics to quantify distributional similarity to the LDCT target domain. The FID score measures the similarity between the feature distributions of the real LDCT images and the generated LDCT images. The lower the FID score, the closer the generated images are to the real data distribution. The KID score uses the maximum mean discrepancy (MMD) to measure the distance between feature distributions. Similar to FID, the lower the KID score, the better the model's performance in generating LDCT images.

\begin{table}[htbp]
    \centering
    \caption{Number of subjects and images from the LIDC dataset used in the training and testing cyclegan.}
    \label{tab:table_2}
    \begin{tabular}{l|cc}
        LIDC-IDRI dataset for CycleGAN training & \textbf{Subjects} & \textbf{Images} \\
        \midrule
        \textbf{Training X – SDCT (80\%)}                    & 556  & 139301 \\
        \textbf{Training Y – LDCT}                         & 315 & 68302 \\
        \textbf{Test X – SDCT (20\%)} & 139 & 35280 \\
    \end{tabular}
\end{table}

\subsubsection{Radiomic pipeline}
\begin{figure} 
    \centering
    \includegraphics[width=0.8\linewidth]{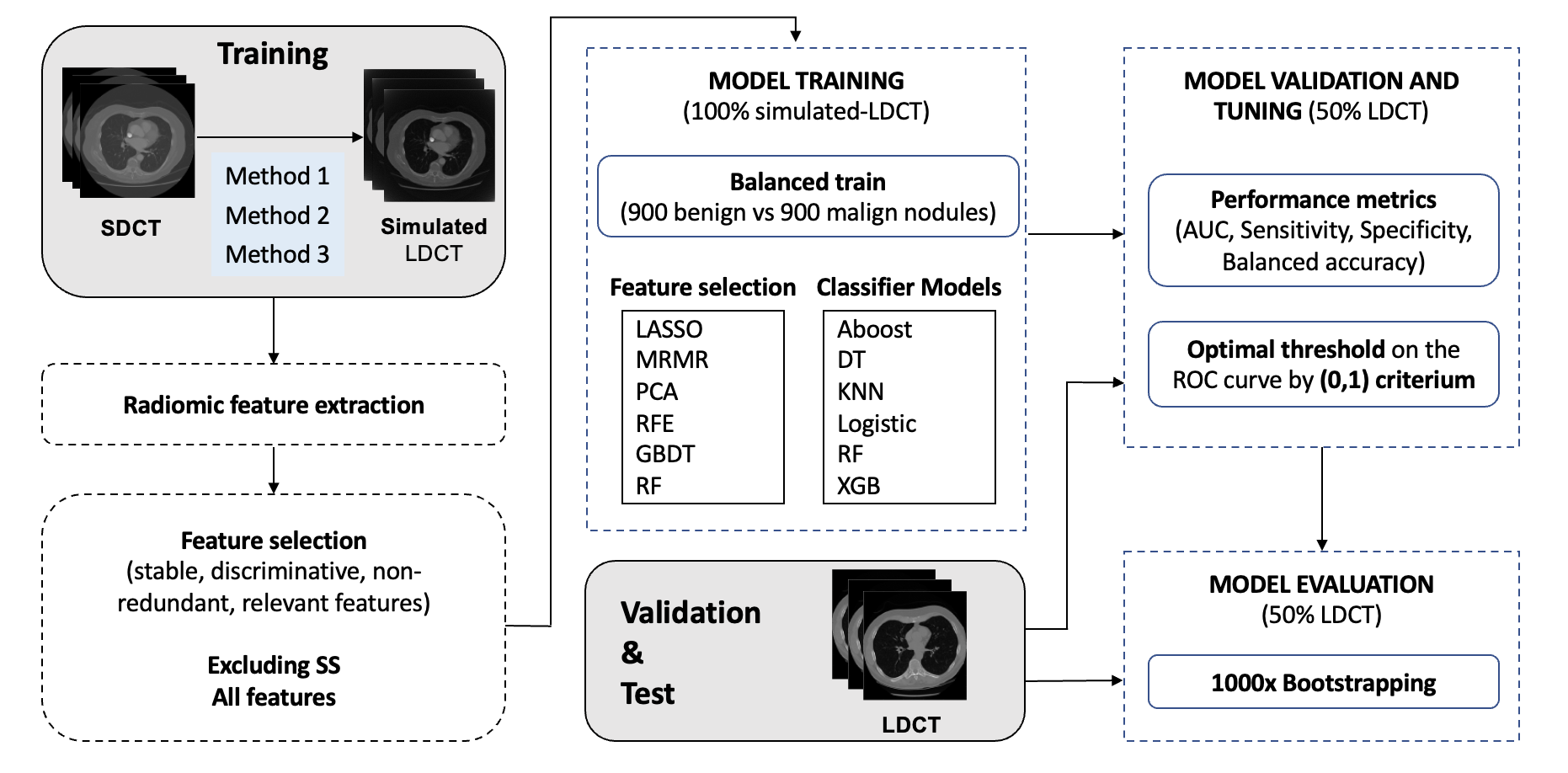}
    \caption{Pipeline for pulmonary nodule classification using radiomics. }
  \label{fig:fig7}
\end{figure}

The three degradation methods were employed as image generator for training a pulmonary nodule classification, following the methodology proposed in our previous work \cite{Liu2024}. This classification task ensures a comparative assessment of each simulation method’s ability to preserve diagnostically relevant information and enhance classification performance. The pipeline is summarized in Fig. \ref{fig:fig7} and consists of (1) radiomic feature extraction, feature selection, (2) balancing the training set, (3) model training, (4) model validation and (5) evaluation including explainability analysis. Radiomic features were extracted using PyRadiomics \cite{vanGriethuysen2017}, following the methodology of our prior work \cite{Liu2024} with settings adapted for normalized CT volumes. Extraction was performed in 3D on isotropically resampled volumes (1×1×1 mm, sitkBSpline interpolation). Intensity values were Z-score normalized prior to extraction, and gray-level discretization used a fixed bin count of 32. Features were computed on both the original image and wavelet-decomposed images, spanning seven feature classes: shape, first-order, GLCM, GLRLM, GLSZM, NGTDM, and GLDM, yielding 851 features in total. PyRadiomics implements features in accordance with the Image Biomarker Standardisation Initiative (IBSI) guidelines, ensuring reproducibility across implementations. The YAML configuration file used for extraction is provided as supplementary material.  Feature selection step included stability and discriminative analysis, the removal of redundant features and the feature selection using absolute shrinkage and selection operator (LASSO), minimum redundancy maximum relevance (MRMR), principal component analysis (PCA), recursive feature elimination (RFE), feature selection using gradient-boosted decision tree (GBDT) and random forest (RF). To address the imbalanced training dataset (1725 non-malignant nodules and 225 malignant nodules), a data augmentation technique introduced in \cite{LoIacono2024} was first applied to the minority class (225 malignant nodules). This augmentation technique consisted in the application of ROI perturbations, including $15\%$ dilation and erosion, as well as random contour adjustments, resulting in a total of 900 malignant nodules. Subsequently, random undersampling was applied to the majority class (1725 benign nodules) to achieve 900 benign nodules. Model training included the training of six classification models: AdaBoost, Decision Tree (DT), k-Nearest Neighbors (KNN), Logistic Regression, Random Forest (RF), and eXtreme Gradient Boosting (XGB). 

During the validation phase, 50\% of the LDCT cases was used to optimize model parameters, including: (1) selecting the optimal decision threshold for the receiver operating characteristic (ROC) curve determined by the minimum Euclidean distance to point (0,1); (2) identifying the optimal feature subset size from 5, 10, 15, 20 features (except for LASSO) based on the area under the ROC curve (AUC); and (3) determining the best feature selection and classifier combination based on a metric obtained by averaging the AUC, balanced accuracy, sensitivity and specificity. The final selected model and feature set for each degradation method were tested on the reserved test set (the remaining 50\% of the real LDCT cases). Finally, during testing, 1000 bootstrapping iterations with 10\% replacement were computed to obtain the average and 95\% confidence interval (CI) for each performance metric. 

To ensure experimental validity, we emphasize the strict separation between the image synthesis and classification pipelines. CycleGAN was trained in the SDCT to LDCT direction using real SDCT images as source and real LDCT images as unpaired target-domain references. The resulting synthetic LDCT images, generated from SDCT subjects, constitute the classifier training set. The 315 real LDCT subjects are used exclusively as the classifier validation and test set and are never or included in classifier training. This design ensures that no test subject contributes, directly or indirectly, to classifier training, and that the evaluation reflects genuine generalization to unseen real LDCT data.

\subsubsection{Statistical analysis for method comparison}
Non-parametric statistical tests were used to compare the performance of the degradation methods. The primary analysis was conducted using the Friedman non-parametric test to determine if there was a statistically significant difference in the performance among the three degradation methods. If the p-value from the Friedman test was less than 0.05, it indicates that the difference is significant and a pairwise comparison is computed using the Wilcoxon signed-rank test with corrected p-values using the Bonferroni correction.  

\section{Results}
This section presents the results of this study, organized in the following way: (1) evaluation of the degradation methods based on visual and quantitative metrics; (2) performance of the radiomics-based pulmonary nodule classification model when trained on the simulated LDCT data; and (3) statistical comparison of the three degradation approaches.

\subsection{Degradation method evaluation}
In this section, we report the intrinsic performance metrics used during model development, as well as the final cross-method qualitative and quantitative comparison. 

\subsubsection{Intrinsic results during model development}

Due to the distinct architectural nature of each degradation method (deterministic, paired and unpaired), different validation metrics were required for optimization prior to the final comparative analysis.

For method 1, as it was a deterministic noise insertion method rather than a learnable parameter-based model, no optimization metrics were applicable. 

For method 2, Pix2Pix paired translation model, performance was evaluated using pixel-wise reconstruction metrics. After 200 epochs, the model achieved a MAE of 0.0961 and a MS-SSIM of 0.7192 on the validation. 

For method 3, the CycleGAN unpaired translation model was evaluated by translating the reserved SDCT test set to the LDCT domain and comparing it against the real LDCT reference distribution. The training performance over epochs is presented in Table \ref{tab:table_3}. The model consistently improved until epoch 20 achieving the lowest FID and KID scores and was thus selected to degrade the SDCT images from the LIDC-IDRI dataset. 

\begin{table}[htbp]
    \centering
    \caption{Results obtained from cyclegan after training for 5, 10, 15, 20 epochs and 21 epochs.}
    \label{tab:table_3}
    \begin{tabular}{l|cc}
        \textbf{Testing epoch} & \textbf{FID} & \textbf{KID (mean$\pm$ std)} \\
        \midrule
        Epoch 5                  & 1.0458  & 2.9562; 1.7077 \\
        Epoch 10                          & 0.3505 & 0.5087$\pm$ 0.5918 \\
        Epoch 15                          & 0.1645 & 0.4115$\pm$ 0.7622 \\
        \textbf{Epoch 20} & \textbf{0.0925} & \textbf{0.1304$\pm$ 0.5235} \\
        Epoch 25                          & 0.1891 & 0.2578$\pm$ 0.4185
    \end{tabular}
\end{table}

\subsubsection{Cross-method qualitative and quantitative comparison}
\begin{figure} 
    \centering
    \includegraphics[width=0.6\linewidth]{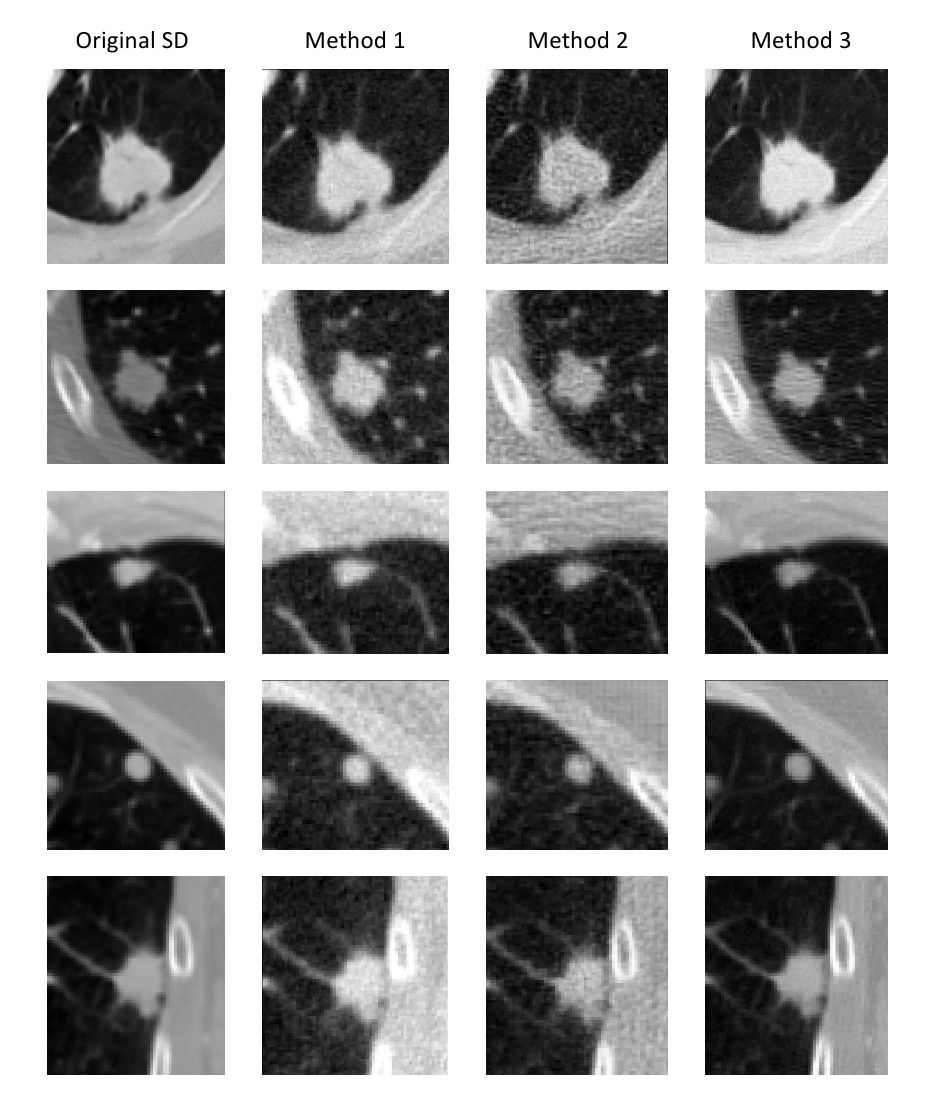}
    \caption{Visual analysis of nodule patches from degradation methods. The figure compares patches from the original SDCT dataset with the corresponding simulated images generated by method 1 (simple sinogram-domain simulation), method 2 (physics-based degradation model replication using Pix2Pix), and method 3 (CycleGAN).}
  \label{fig:fig8}
\end{figure}

The visual characteristics of the simulated LDCT patches in comparison to the original SDCT patch are presented in Fig.\ref{fig:fig8}. All three methods introduce visible non-uniform noise compared to the original SD patch. Method 1 presents a noticeable fine granular noise, similar to salt-and-pepper noise, while method 2 presents more pronounced, coarse-grained noise with distinct streak artifacts visible in high-density regions, such as chest wall. Method 3 shows blurriness, a characteristic found in LDCT images.

To provide a quantitative comparison, distributional similarity metrics FID and KID scores were computed using 128x128 center-cropped CT slice patches from degraded images, referencing the real LDCT slices as the target ground truth distribution (see Table \ref{tab:table_4}).

The baseline SDCT images obtained an FID of 1.1494 and mean KID of 1.3695. Notably, method 1 and method 2 obtained a higher FID and KID scores compared to the SDCT baseline (method 1 FID: 4.0717; method 2 FID: 4.6729), indicating that while visually these methods introduced noise artifacts, the statistical characteristics of this synthetic noise deviated from the noise of real LDCTs, causing a divergence in the deep feature space. In contrast, method 3 outperformed both the baseline and the paired degradation methods, achieving the lowest scores (FID: 0.1734; KID: 0.0813). This means method 3 effectively translated SDCT images into a distribution that statistically aligns with the real LDCT target domain. This superior performance is consistent with the validation strategy used for the CycleGAN framework, where FID and KID metrics were utilized as the selection criteria to identify the optimal training checkpoint that maximized distributional alignment.

\begin{table}[htbp]
    \centering
    \caption{Fid and kid scores computed on image patches. Lower scores represent better distributional alignment with the real ldct domain.}
    \label{tab:table_4}
    \begin{tabular}{l|cc}
        & \textbf{FID} & \textbf{KID (mean$\pm$std)} \\
        \midrule
        \textbf{SDCT}                 & 1.1494  & 1.3695$\pm$ 0.5763 \\
        \textbf{Method 1}             & 4.0717	& 3.0200 $\pm$ 0.5110  \\
        \textbf{Method 2}             & 4.6729	& 8.4469 $\pm$ 1.1596 \\
        \textbf{Method 3} & \textbf{0.1734} & \textbf{0.0813 $\pm$ 0.1002} \\
    \end{tabular}
\end{table}
\subsection{Radiomics-based pulmonary nodule classification}
Prior to model training, radiomic feature analysis including stability, discriminative power, and redundancy removal, was performed on the features extracted from the nodules from each simulated dataset to obtain a smaller set of robust radiomic features \cite{Liu2024}.

Fig. 9 illustrated the outcome of the feature analysis process across the three degradation methods and the SDCT baseline. Despite the varying noise characteristics introduced by each simulation technique, as described before and shown in Fig.8, the number of stable, discriminative and non-redundant features identified was consistent among the degradation methods: 99 from methods 1 and 2, and 103 from method 3. In comparison, the SDCT baseline identified 116 features. Table \ref{tab:table_5} provides a categorical feature count of the final stable, discriminative and non-redundant feature sets, indicating wavelet-derived features as the most selected across all methods. The intersection ($\cap$) columns highlight the cross-method overlap: 60 features were found in common considering the three degradation methods (m1 $\cap$ m2 $\cap$ m3), while 52 of these features were also present in the SDCT baseline (SDCT$\cap$ m1$\cap$m2$\cap$m3), representing the features robust to both degradation and domain shifts. See supplementary materials for detailed feature list.

\begin{figure} 
    \centering
    \includegraphics[width=0.6\linewidth]{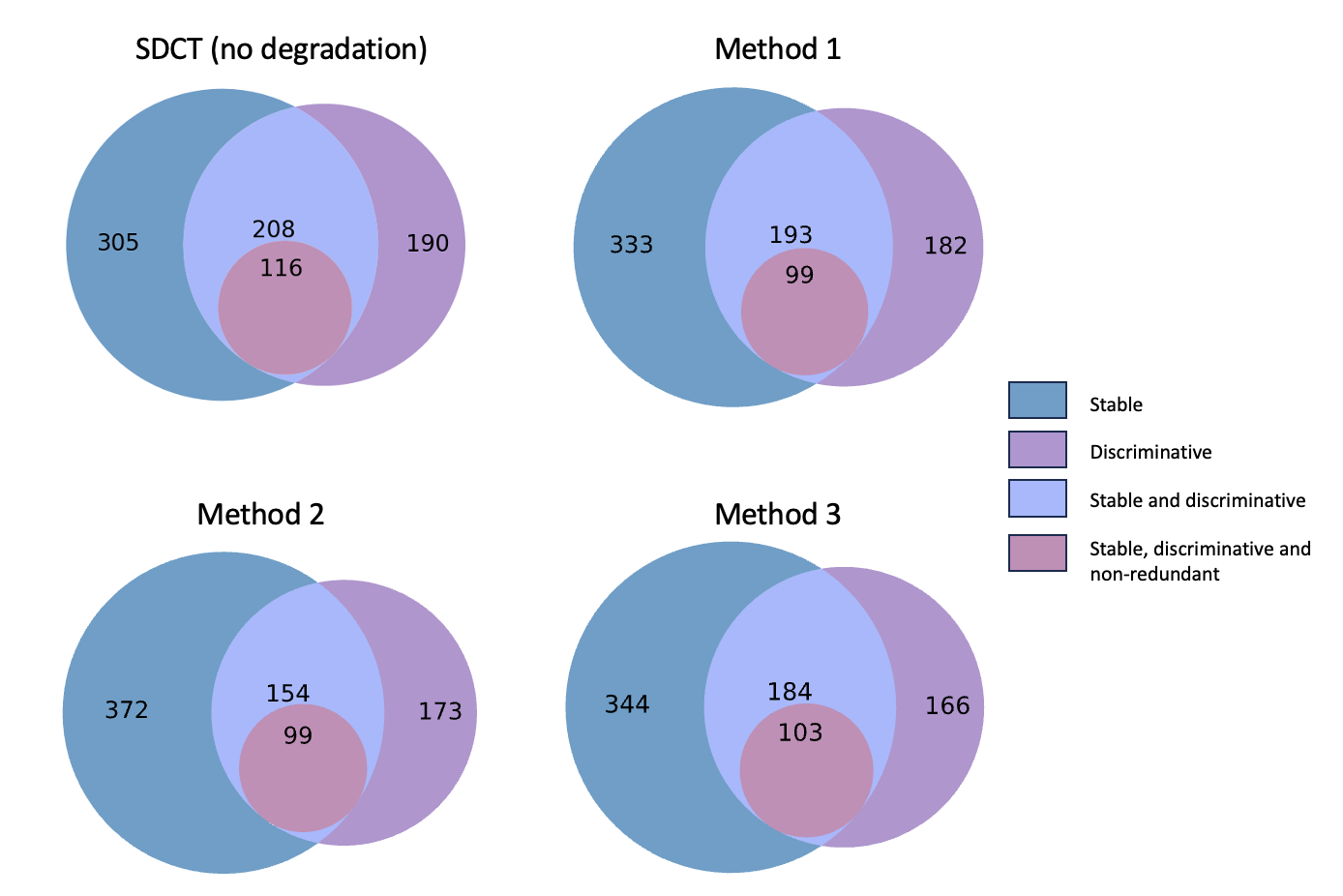}
    \caption{Venn diagram representing the number of features considered stable, discriminant and non-redundant for each method.}
  \label{fig:fig9}
\end{figure}

\begin{table}[htbp]
    \centering
    \caption{Feature counts by category and cross-method intersection analysis. The "m1 $\cap$ m2 $\cap$ m3" column represents features present in all three degradation methods, while the final column includes the SDCT baseline in the intersection.}
    \label{tab:table_5}
    \begin{tabular}{l|cccc|cc}
        & \textbf{m1} & \textbf{m2} & \textbf{m3} & \textbf{m1$\cap$m2$\cap$m3} & \textbf{SDCT} & \textbf{SDCT$\cap$m1$\cap$m2$\cap$m3} \\
        \midrule
        \textbf{original shape}    & 2 & 2 & 2 & 2 & 2 &	2 \\
        \textbf{original first order}     &7 &	7 & 7 &	5 &	5 &	4\\
        \textbf{original textural}         & 9 & 9 & 9 & 5 & 7 & 3 \\
        \textbf{wavelet} & 81	& 81	& 85& 	48	& 102& 	43 \\
        \midrule
        \textbf{Total} & 99& 	99& 	103& 	60& 	116	& 52 \\
    \end{tabular}
\end{table}

The optimal feature set subset size (5, 10, 15, 20) and the best pair of feature selector and classifier combination are reported in the supplementary material. The validation and test results for each degradation scenario are summarized in Table \ref{tab:tabl_6_classification}. Performance metrics for the test set are reported as average and 95\% CI from 1000 bootstrapped samples.
The overall classification performance on the real LDCT test set highlights the importance of degrading the SDCT images for domain-specific data augmentation. The model trained on baseline SDCT (without degradation) performed well during validation (AUC 0.863) but decreased on the testing phase (AUC 0.789). Specifically, the sensitivity decreased to 0.571 which indicates that the model failed to classify the nodules from the LDCT test set. A detailed statistical comparison of these results is presented in the following section.
On the other hand, all three degradation methods effectively simulated LDCT features to train a robust nodule classification model, achieving a test AUC above 0.84 for all the methods and settings. From validation to testing, the AUC metric remained robust across all method. Method 1 maintained an identical AUC (0.844) on both validation and test, while method 2 and method 3 improved slightly by 1.6\% and 0.7\%, respectively. Conversely, balanced accuracy showed general decrease across all methods, with method 1 obtaining a decrease of 4.6\%, method 2 of 3.3\%, and method 3 showing the most stability with only a reduction of only 1.4\%.  In terms of sensitivity, method 1 and 2 showed a notable decrease of 10.9\% and 8.1\%, respectively, in the test set. This decrease was compensated by high specificity, increasing 1.7\% for method 1 and 1.3\% for method 2. Method 3 demonstrated the most balanced performance on the independent test set, maintaining the highest sensitivity (0.743) and balanced accuracy (0.800), with increases in AUC (0.7\%) and specificity (5.3\%) from validation to test. This suggests that the unpaired CycleGAN simulation (method 3) may preserve subtle nodule features more effectively than the paired approaches, reducing the rate of false negatives. 

\begin{table}[htbp]
    \centering
    \caption{Nodule classification validation and test results.}
    \label{tab:tabl_6_classification}
    \begin{tabular}{lcccc}
        \toprule
         & \textbf{AUC} & \textbf{Balanced accuracy} & \textbf{Sensitivity} & \textbf{Specificity} \\
        \midrule

        \textbf{SDCT} & & & & \\
        \quad validation & 0.863 & 0.814 & 0.824 & 0.805 \\
        \quad test & 0.789 {[}0.750; 0.827{]} & 0.744 {[}0.707; 0.781{]} & 0.571 {[}0.500; 0.643{]} & 0.917 {[}0.904; 0.930{]} \\
        \midrule

        \textbf{Method 1} & & & & \\
        \quad validation & 0.844 & 0.813 & 0.765 & 0.861 \\
        \quad test & 0.844 {[}0.804; 0.886{]} & 0.767 {[}0.731; 0.803{]} & 0.656 {[}0.586; 0.727{]} & 0.878 {[}0.862; 0.893{]} \\
        \midrule

        \textbf{Method 2} & & & & \\
        \quad validation & 0.843 & 0.821 & 0.765 & 0.878 \\
        \quad test & 0.859 {[}0.824; 0.900{]} & 0.788 {[}0.752; 0.822{]} & 0.684 {[}0.618; 0.757{]} & \textbf{0.891 {[}0.877; 0.905{]}} \\
        \midrule

        \textbf{Method 3} & & & & \\
        \quad validation & 0.854 & 0.814 & 0.824 & 0.805 \\
        \quad test & \textbf{0.861 {[}0.826; 0.900{]}} & \textbf{0.800 {[}0.768; 0.836{]}} & \textbf{0.743 {[}0.684; 0.811{]}} & 0.858 {[}0.842; 0.875{]} \\
        \bottomrule
    \end{tabular}
\end{table}

\subsection{Statistical test for degradation methods comparison}
The performance of the three degradation methods was formally compared using the 1000 bootstrapped test set AUC, balanced accuracy, sensitivity and specificity values.

For every metric, the Friedman test consistently yielded a p-value<0.05 (see supplementary materials) when comparing the three degradation methods, indicating that at least one method's performance distribution was statistically significantly different from the others. Subsequent Bonferroni-corrected pairwise Wilcoxon signed-rank tests confirmed these differences, reporting p-value<0.05 for all metrics (see supplementary materials).

Fig. 10 represents bootstrapped distribution violin plots, highlighting a distinct domain shift between the SDCT baseline and the degradation methods. The SDCT baseline showed lower distribution across most metrics compared to the degradation approaches. More specifically, the SDCT baseline presented a drop in sensitivity distribution (0.571 [0.500; 0.643]), indicating a failure to classify nodules in the noisy LDCT domain.

\begin{figure} 
    \centering
    \includegraphics[width=0.8\linewidth]{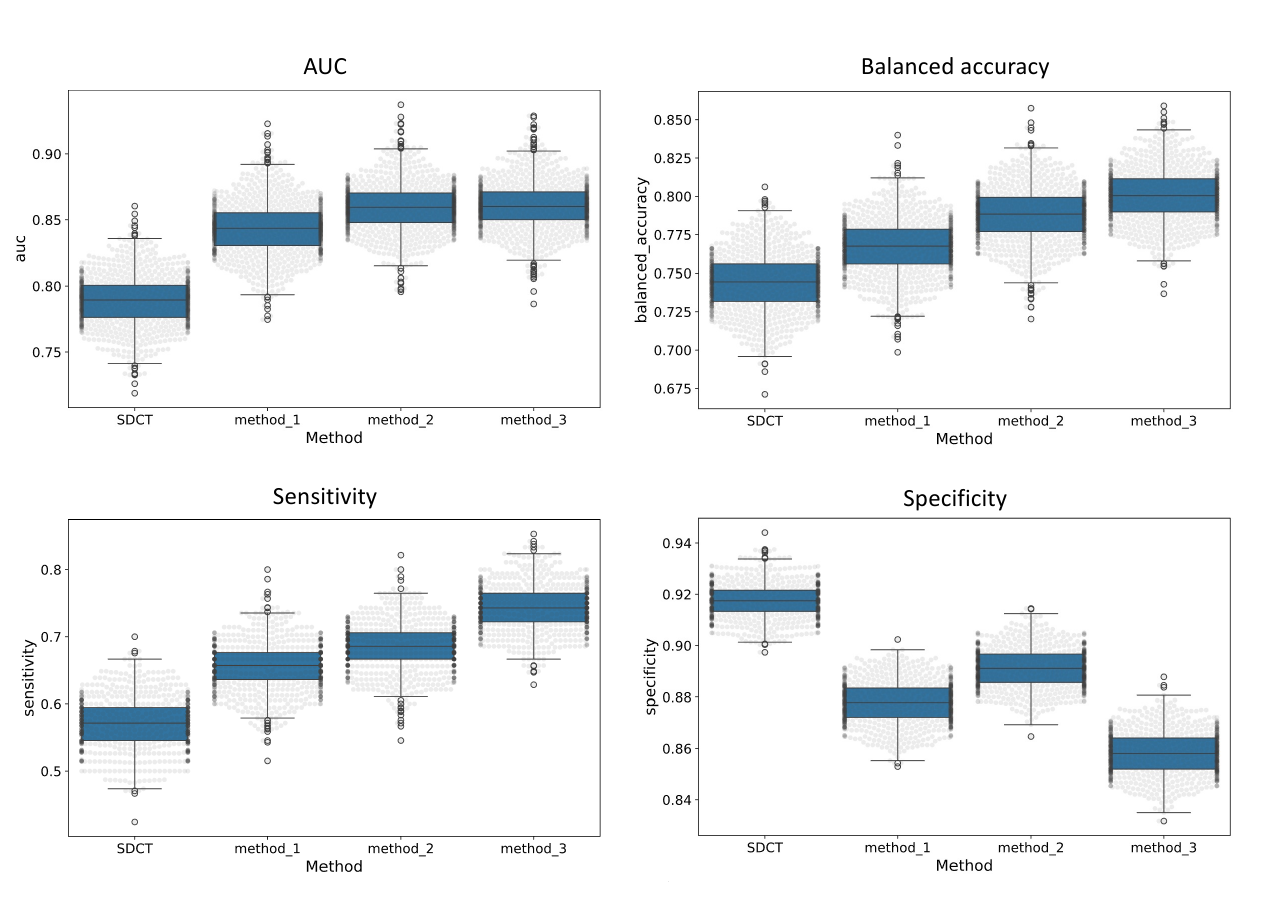}
    \caption{Bootstrapped distribution violin plots of the nodule classification model on the independent real LDCT test set. The boxplots illustrate the variability across 1000 bootstrap iterations for AUC, balanced accuracy, sensitivity and specificity. }
  \label{fig:fig10}
\end{figure}
Focusing on comparing the degradation methods, AUC distributions for methods 2 and 3 are visually similar distributions with highly overlapping confidence intervals (method 2: 0.859 [0.824; 0.900] and method 3: 0.861 [0.826; 0.900]), both slightly superior to method 1 (0.844 [0.804; 0.886]). In terms of balanced accuracy, a sequential increase is observed from method 1 to method 3, with method 3 achieving the highest results: method 3 (0.800 [0.768; 0.836] > method 2 (0.788 [0.752; 0.822]) > method 1 (0.767 [0.731; 0.803]). Sensitivity and specificity reveal a distinct performance trade-off as well as a difference in variability. A stepwise increase is also observed in sensitivity moving from method 1 to method 3, where method 3 achieves the highest median sensitivity of 0.743 [0.684; 0.811]. However, compared to the other metrics, sensitivity exhibits the widest confidence intervals, indicating greater variability in the model’s ability to detect positive cases. Conversely, the specificity plot shows the inverse trend characterized by significantly narrower confidence intervals, reflecting higher stability. Method 2 dominates with the highest specificity (0.891 [0.877; 0.905]), while Method 3 shows a lower, albeit still robust, specificity distribution (0.858 [0.842; 0.875]). 
Despite the partial overlap in the 95\% confidence intervals, the statistical analysis confirms that the underlying performance distributions are distinct. The consistent shift in the distributions confirms that method 3 effectively prioritizes the detection of true positives (sensitivity) compared to methods 1 and 2.

\section{Discussion}

This study provided a systematic comparison of different LDCT degradation methods for synthetic image generation and evaluated the efficacy in improving nodule classification models using radiomic features extracted from those synthetic images. We included approaches ranging from sinogram-domain statistical simulation (method 1), to the replication of a validated physics-based method using Pix2Pix (method 2), to unpaired image generation using CycleGAN (method 3). In this way, we bypassed the conventional denoising step and instead validated the feasibility of using LDCT images to develop downstream classification models.
The main contributions of this paper are:

\begin{itemize}
\item We provide the first systematic comparison and evaluation of different low-dose CT degradation methods.
\item We introduce an approach that bypasses denoising and instead validates the feasibility of using SDCT images for developing downstream classification models.
\item By providing a reliable method for simulating LDCT scans, our findings will not only aid in the development of more accurate denoising algorithms but also accelerate the inclusion of SDCT images into the development of nodule detection and classification models, ultimately improving the efficacy of lung cancer screening.
\end{itemize}

The need for realistic degradation was confirmed by the poor generalization of the model trained on SDCT data, i.e., images without any degradation. While the SDCT baseline model performed well during validation, it failed to generalize to the test set, yielding a collapsed sensitivity of 0.571. In contrast, the results demonstrated that synthetic data generated by all three methods can successfully train a classifier to perform well on real LDCT data (AUC > 0.84). However, the methods showed distinct qualitative visual features and distinct quantitative performance in the nodule classification task. Method 1 produced visible but more subtle granular noise compared to method 2, which replicated pronounced streak artifacts and high-frequency noise that appeared visually consistent with the photon starvation. However, despite visible noise in methods 2, it yielded relatively lower sensitivity. Conversely, method 3 produced perceptually blurrier images with subtle noise but obtained the best FID (0.0316) and KID (0.0102) scores computed on nodule. This indicates that while method 3 may lack high-frequency sharpness, it achieved superior distributional alignment with the target domain. In the nodule classification task, method 3 yielded the most balanced classification performance using with the highest sensitivity, achieving a test set AUC of 0.861, balanced accuracy of 0.800, sensitivity of 0.743, specificity of 0.858.
In terms of radiomic feature analysis, despite the visual differences in the images generated by different methods, the final features identified as stable, discriminant and non-redundant were similar, with 60 features found in common. The dominance of wavelet-derived features (80\% of this common set), specifically those with low-pass components (92 low-pass directional filter components compared to 52 high-pass components), indicates that the underlying structural characteristics of the nodules are preserved regardless of the degradation technique, while high-frequency noise variations are less relevant for discriminating features. When considering the feature subsets selected to train the classifiers, only two features were commonly selected across all methods, original shape major axis length and original shape sphericity. This highlights the importance of these SS features and is coherent with the criteria of the Lung Imaging Reporting and Data System (Lung-RADS) \cite{Christensen2024}, which primarily quantifies tumor risk based on nodule diameter and morphological regularity. These findings are consistent with our previous work \cite{Liu2024}. Further details regarding the selected features can be found in the supplementary materials.  
This study is not exempt from limitations. First, the simulation methods relied on image-domain or simplified statistic assumptions in the sinogram domain, future work should consider scanner specific parameters or leverage deep learning in the projection domain to better include non-uniform realistic LDCT noise. Second, regarding the statistical analysis, although the Friedman and pairwise Wilcoxon tests indicated significant differences, the overlap in the 95\% confidence intervals suggest that these differences might be partially attributed to the bootstrap sample size which makes the statistical tests more susceptible to small differences. Finally, while the classification results are promising, the sensitivity (0.743) of the models could be improved, despite being comparable to previous findings \cite{Liu2024}. However, the high specificity (0.858) is clinically significant, as a major limitation of LDCT screening programs is the high rate of false positives leading to unnecessary interventions. Future work should validate the proposed pipeline on independent annotated LDCT datasets such as LUNA25 to assess generalizability of the model.

\bibliographystyle{unsrt}  
\bibliography{references}  

@article{Sung2021,
  author  = {Sung, H. and others},
  title   = {Global Cancer Statistics 2020: {GLOBOCAN} Estimates of Incidence and Mortality Worldwide for 36 Cancers in 185 Countries},
  journal = {CA: A Cancer Journal for Clinicians},
  volume  = {71},
  number  = {3},
  pages   = {209--249},
  month   = may,
  year    = {2021},
  doi     = {10.3322/caac.21660}
}

@article{deKoning2020,
  author  = {de Koning, H. J. and others},
  title   = {Reduced Lung-Cancer Mortality with Volume {CT} Screening in a Randomized Trial},
  journal = {New England Journal of Medicine},
  volume  = {382},
  number  = {6},
  pages   = {503--513},
  month   = feb,
  year    = {2020},
  doi     = {10.1056/NEJMoa1911793}
}

@article{NLST2011,
  author  = {{The National Lung Screening Trial Research Team}},
  title   = {Reduced Lung-Cancer Mortality with Low-Dose Computed Tomographic Screening},
  journal = {New England Journal of Medicine},
  volume  = {365},
  number  = {5},
  pages   = {395--409},
  month   = aug,
  year    = {2011},
  doi     = {10.1056/NEJMoa1102873}
}

@article{Kim2024,
  author  = {Kim, W. and Jeon, S. Y. and Byun, G. and Yoo, H. and Choi, J. H.},
  title   = {A systematic review of deep learning-based denoising for low-dose computed tomography from a perceptual quality perspective},
  journal = {Biomedical Engineering Letters},
  volume  = {14},
  number  = {6},
  pages   = {1153--1173},
  month   = aug,
  year    = {2024},
  doi     = {10.1007/S13534-024-00419-7}
}

@article{Kulathilake2023,
  author  = {Kulathilake, K. A. S. H. and Abdullah, N. A. and Sabri, A. Q. M. and Lai, K. W.},
  title   = {A review on Deep Learning approaches for low-dose Computed Tomography restoration},
  journal = {Complex \& Intelligent Systems},
  volume  = {9},
  number  = {3},
  pages   = {2713--2745},
  month   = jun,
  year    = {2023},
  doi     = {10.1007/S40747-021-00405-X}
}

@article{Zhao2019,
  author  = {Zhao, T. and McNitt-Gray, M. and Ruan, D.},
  title   = {A convolutional neural network for ultra-low-dose {CT} denoising and emphysema screening},
  journal = {Medical Physics},
  volume  = {46},
  number  = {9},
  pages   = {3941--3950},
  month   = sep,
  year    = {2019},
  doi     = {10.1002/MP.13666}
}

@article{Gholizadeh2019,
  author  = {Gholizadeh-Ansari, M. and Alirezaie, J. and Babyn, P.},
  title   = {Deep Learning for Low-Dose {CT} Denoising Using Perceptual Loss and Edge Detection Layer},
  journal = {Journal of Digital Imaging},
  volume  = {33},
  number  = {2},
  pages   = {504--515},
  month   = sep,
  year    = {2019},
  doi     = {10.1007/S10278-019-00274-4}
}

@inproceedings{Chen2017,
  author    = {Chen, H. and others},
  title     = {Low-dose {CT} denoising with convolutional neural network},
  booktitle = {Proceedings of the International Symposium on Biomedical Imaging (ISBI)},
  pages     = {143--146},
  month     = jun,
  year      = {2017},
  doi       = {10.1109/ISBI.2017.7950488}
}

@article{Eulig2024,
  author  = {Eulig, E. and Ommer, B. and Kachelrie{\ss}, M.},
  title   = {Benchmarking deep learning-based low-dose {CT} image denoising algorithms},
  journal = {Medical Physics},
  volume  = {51},
  number  = {12},
  pages   = {8776--8788},
  month   = dec,
  year    = {2024},
  doi     = {10.1002/MP.17379}
}

@article{McCollough2017,
  author  = {McCollough, C. H. and others},
  title   = {Low-dose {CT} for the detection and classification of metastatic liver lesions: Results of the 2016 Low Dose {CT} Grand Challenge},
  journal = {Medical Physics},
  volume  = {44},
  number  = {10},
  pages   = {e339--e352},
  month   = oct,
  year    = {2017},
  doi     = {10.1002/MP.12345}
}

@article{Armato2011,
  author  = {Armato, S. G. and others},
  title   = {The Lung Image Database Consortium ({LIDC}) and Image Database Resource Initiative ({IDRI}): A completed reference database of lung nodules on {CT} scans},
  journal = {Medical Physics},
  volume  = {38},
  number  = {2},
  year    = {2011},
  doi     = {10.1118/1.3528204}
}

@article{Moen2021,
  author  = {Moen, T. R. and others},
  title   = {Low-dose {CT} image and projection dataset},
  journal = {Medical Physics},
  volume  = {48},
  number  = {2},
  pages   = {902--911},
  month   = feb,
  year    = {2021},
  doi     = {10.1002/MP.14594}
}

@article{Snowsill2018,
  author  = {Snowsill, T. and others},
  title   = {Low-dose computed tomography for lung cancer screening in high-risk populations: a systematic review and economic evaluation},
  journal = {Health Technology Assessment},
  volume  = {22},
  number  = {69},
  pages   = {1--276},
  month   = nov,
  year    = {2018},
  doi     = {10.3310/HTA22690}
}

@article{Rampinelli2013,
  author  = {Rampinelli, C. and Origgi, D. and Bellomi, M.},
  title   = {Low-dose {CT}: technique, reading methods and image interpretation},
  journal = {Cancer Imaging},
  volume  = {12},
  number  = {3},
  pages   = {548},
  year    = {2013},
  doi     = {10.1102/1470-7330.2012.0049}
}

@article{Yu2012,
  author  = {Yu, L. and Shiung, M. and Jondal, D. and McCollough, C. H.},
  title   = {Development and validation of a practical lower-dose-simulation tool for optimizing computed tomography scan protocols},
  journal = {Journal of Computer Assisted Tomography},
  volume  = {36},
  number  = {4},
  pages   = {477--487},
  year    = {2012},
  doi     = {10.1097/RCT.0B013E318258E891}
}

@article{Zeng2015,
  author  = {Zeng, D. and others},
  title   = {A simple low-dose {X}-Ray {CT} simulation from high-dose scan},
  journal = {IEEE Transactions on Nuclear Science},
  volume  = {62},
  number  = {5},
  pages   = {2226--2233},
  month   = oct,
  year    = {2015},
  doi     = {10.1109/TNS.2015.2467219}
}

@inproceedings{Isola2017,
  author    = {Isola, P. and Zhu, J. Y. and Zhou, T. and Efros, A. A.},
  title     = {Image-to-image translation with conditional adversarial networks},
  booktitle = {Proceedings of the IEEE Conference on Computer Vision and Pattern Recognition (CVPR)},
  year      = {2017},
  doi       = {10.1109/CVPR.2017.632}
}

@inproceedings{Zhu2017,
  author    = {Zhu, J. Y. and Park, T. and Isola, P. and Efros, A. A.},
  title     = {Unpaired Image-to-Image Translation using Cycle-Consistent Adversarial Networks},
  booktitle = {Proceedings of the IEEE International Conference on Computer Vision (ICCV)},
  pages     = {2242--2251},
  month     = mar,
  year      = {2017},
  doi       = {10.1109/ICCV.2017.244}
}

@article{Liu2024,
  author  = {Liu, J. and Corti, A. and Corino, V. D. A. and Mainardi, L.},
  title   = {Lung nodule classification using radiomics model trained on degraded {SDCT} images},
  journal = {Computer Methods and Programs in Biomedicine},
  volume  = {257},
  pages   = {108474},
  month   = dec,
  year    = {2024},
  doi     = {10.1016/J.CMPB.2024.108474}
}

@article{Christensen2024,
  author  = {Christensen, J. and others},
  title   = {{ACR} Lung-{RADS} v2022: Assessment Categories and Management Recommendations},
  journal = {Chest},
  volume  = {165},
  number  = {3},
  pages   = {738--753},
  month   = mar,
  year    = {2024},
  doi     = {10.1016/j.chest.2023.10.028}
}

@article{vanGriethuysen2017,
  author  = {van Griethuysen, J. J. M. and others},
  title   = {Computational Radiomics System to Decode the Radiographic Phenotype},
  journal = {Cancer Research},
  volume  = {77},
  number  = {21},
  pages   = {e104--e107},
  month   = nov,
  year    = {2017},
  doi     = {10.1158/0008-5472.CAN-17-0339}
}

@article{LoIacono2024,
  author  = {Lo Iacono, F. and Maragna, R. and Pontone, G. and Corino, V. D. A.},
  title   = {A Novel Data Augmentation Method for Radiomics Analysis Using Image Perturbations},
  journal = {Journal of Imaging Informatics in Medicine},
  volume  = {37},
  number  = {5},
  pages   = {2401--2414},
  month   = may,
  year    = {2024},
  doi     = {10.1007/S10278-024-01013-0}
}






\end{document}